\documentclass[aps,pra,showpacs,groupedaddress]{revtex4}
\usepackage{amssymb,bm}
\usepackage[dvips]{graphicx}
\usepackage{graphicx}
\usepackage{dcolumn}

\begin{document}
\bibliographystyle{apsrev}

\title{\hfill {\footnotesize FZJ--IKP(TH)--2005--23}\\
Spin observables of the reactions $NN\to \Delta N$ and
 $pd\to \Delta (pp)(^1S_0)$ in collinear kinematics}


\author{Yu.N.~Uzikov} 
\email{uzikov@nusun.jinr.ru}
\affiliation{ Joint Institute for Nuclear Research, LNP, 141980 Dubna,
 Moscow Region, Russia}

\author{J.~Haidenbauer}
\email{j.haidenbauer@fz-juelich.de}
\affiliation{ Institut f\"ur Kernphysik, Forschungszentrum J\"ulich GmbH, 
D-52425 J\"ulich, Germany} 
\date{\today}
\begin{abstract}
{A general formalism for double and triple spin-correlations of the reaction
 ${\vec N}{\vec N}\to {\vec \Delta}N$  is developed 
for the case of collinear kinematics.
 A complete polarization experiment allowing to reconstruct all of the four 
 amplitudes describing  this process is suggested. Furthermore, the spin 
 observables of the inelastic charge-exchange reaction 
 ${\vec p}{\vec d}\to {\vec {\Delta^0}}(pp)(^1S_0)$
 are analyzed in collinear kinematics
 within the single $pN$ scattering mechanism involving the subprocess
 ${ p}{ n}\to { \Delta^0}p$. The full set of spin 
 observables related to the polarization of one or two
 initial particles and one final particle is obtained in terms of three
 invariant amplitudes of the reaction 
 $pd\to \Delta(pp)(^1S_0)$ and the transition form factor $d\to (pp)(^1S_0)$.
 A complete polarization experiment for the reaction 
 ${\vec p}{\vec d} \to {\vec {\Delta^0}}(pp)(^1S_0)$ 
 is suggested which allows one to determine three independent combinations 
 of the four amplitudes of the elementary
 subprocess  ${\vec N}{\vec N}\to {\vec \Delta}N$.
}
\end{abstract}  



\pacs{13.88.+e, 13.75.Cs, 14.20.-c}
\maketitle

\section{Introduction}

The $\Delta(1232)$ isobar is a well established nucleon resonance with
spin-parity $j^\pi=\frac{3}{2}^+$ and isospin $I=\frac{3}{2}$.
Due to its coupling to the nucleon-nucleon $(NN)$ system via the transitions 
$NN\to \Delta N$ and $NN\to \Delta \Delta$ 
this resonance plays an important role in the dynamics of the 
$NN$ interaction 
\cite{Tjon,Matsuyama,MHE,Elster,haiden,eyser,machlslaus}
as well as in electromagnatic and pionic processes involving the
$NN$ system, cf., e.g., Refs. \cite{Arenhovel,Jetter,carzilazo-mizutani}
and references therein. 
For example,
unpolarized cross sections of $\pi$-meson production 
at energies of several hundred MeV in many reactions involving
the $NN-NN\pi$ system can be reasonably described by using
theoretical models with the $\Delta$-excitation explicitely included
\cite{carzilazo-mizutani}.
But even near the pion-production threshold the contributions from 
$NN\to \Delta N$ transitions are 
essential for a quantitative description of the
reaction $NN\to NN\pi$ \cite{Jouni,Hanhart1,Hanhart2}.
Also
an important part of three-nucleon forces can be related to the
$\Delta$-isobar excitation, as is widely discussed in the
literature, see, for example, Refs. \cite{kolck,sauer1,sauer2,sauer3}.
These forces allow to remove the so-called
Sagara descrepancy \cite{Sagara} in the unpolarized $nd$ elastic cross
section in the region of the minimum of the cross section \cite{sauer4,witala}
at beam energies 60-200 MeV. The contribution of these forces
increases with beam energy, since the intermediate $\Delta$-isobar
comes closer to its mass-shell, and they dominate the unpolarized
cross section of $pd$ elastic backward scattering at 400-600
MeV, as is seen from model calculations \cite{lak,uzikovrev,boudard}.
Finally, the influence of the $\Delta$-excitation was shown for in-medium
three-body forces too where the $\Delta$ generates an repulsive effective
interaction \cite{samarrucca2001,zuo,fuchs}.
 
Despite of this important role one has to concede that the spin 
dependence of the $NN\to \Delta N$ transition amplitude is not yet 
well known. Indeed the $NN\to \Delta N$ amplitude was studied in 
several experimental \cite{wicklund,shypit,bugg}
and theoretical papers \cite{anisovich,hoshizaki,mizutani,sammarruca}.
However, though a rather satisfactory overall description
of available data at intermediate energies could be achieved, 
there are several unresolved problems concerning specific spin observables
in the $NN\to N\Delta$ transition \cite{mizutani}. 
Also, while three-body forces related to the $\Delta$ 
contribution are capable of explaining the unpolarized $ pd\to pd$ 
cross section, corresponding investigations 
of polarization observables show no systematic improvement
when such three-body forces are included \cite{sakai,sekiguchi,yaschenko}.
Improved information on the spin dependence of the $NN\to \Delta N$
amplitude might allow to shed light on this issue too. 
Furthermore,
a better knowledge of the spin dependence of the $NN\to \Delta N$ 
amplitude might be useful for understanding the remarkable variation 
in the energy dependence of the spin-dependend $pp$ cross sections
$\Delta\sigma_T$ \cite{deboer,ditzler,perrot} and $\Delta\sigma_L$ \cite{auer},
at beam energies 600-800 MeV.
This variation is considered by some authors \cite{hoshizakim,bhandari}
as an indication of dibaryon resonances (for a review see \cite{llehar}).
But since the observed structure lies in the proximity of the nominal 
$\Delta N$ threshold it could be closely linked with the
properites of the $NN\to \Delta N$ and/or
$N\Delta$ amplitudes \cite{kloetsilbar,niskanen1982,anisovich}.

The general binary reaction $NN\to\Delta N$ is described by 16 
independent spin amplitudes and the determination of all of them is
an extremely complicated and challenging experimental task.
In collinear kinematics, however, the spin structure of any binary reaction
 is simplified considerably because only a single physical direction 
 exists in this case, namely the direction of the beam momentum. 
 As a consequence, the number of invariant spin amplitudes describing the
 reactions reduces drastically. A similar simplification occurs in the 
 threshold regime with the s-wave dominance in the final state of the 
 reaction. 
 Under these conditions it seems rather realistic to perform a complete 
 polarization experiment for the process $NN\to N\Delta$ 
 and, as a result, provide stringent constraints on the pertinent transition 
 amplitudes. 
  Note that the analysis of polarization effects in collinear
 kinematics
 with explicit axial symmetry cannot be considered as a limiting case
 of the general formalism developed for binary reactions where the scattering
 plane is well defined \cite{rekaloepan}.
 For the case of collinear kinematics the corresponding
 analytical expressions have to be derived anew. For example,
 for backward
 elastic $dp$ scattering the formalism was developed in \cite{rekalodp}.
The near threshold formalism for
 polarization phenomena has been recently documented in Ref. \cite{rekaloepan}.
 
 The main goal of this paper is to derive the formalism for spin
 observables of binary reactions of the types
 $\frac{1}{2}+\frac{1}{2}\to \frac{3}{2}+\frac{1}{2}$ and
 $\frac{1}{2}+1\to \frac{3}{2}+0$ in collinear kinematics and to 
 find on this basis the corresponding complete polarization experiments
 which would allow one to measure all invariant amplitude describing
 these processes. For this aim we use the technique of Clebsch-Gordan
 coefficients \cite{bilenky} applied recently for collinear kinematics
 of a binary reaction in Ref. \cite{uzikov2} (see for details also Ref.
\cite{uzikov3}).
 Our phenomenological 
analysis of the reactions $NN\to \Delta N$ and $pd\to \Delta (pp)(^1S_0)$
 is  model-independent and  based  only on parity and angular
 momentum conservation. However, in addition, for the reaction 
 $pd\to \Delta (pp)(^1S_0)$
 we develop also a model within the impulse approximation. This allows
 us to connect the three invariant spin amplitudes of the reaction 
 $pd\to \Delta (pp)(^1S_0)$ with the four
 invariant  amplitudes of the process $NN\to \Delta N$. As a result,
 the knowledge of all three independent amplitudes of the reaction
 $pd\to \Delta (pp)(^1S_0)$ obtained from a dedicated complete
 polarizations experiment for this reaction allows one to determine
 the three independent combinations of the four amplitudes of the reaction 
 $NN\to \Delta N$. Earlier, in Ref.~\cite{buggcw} a similar formalism
 was developed for the charge-exchange reaction $dp\to (pp)n$.
 This formalism was succesfully applied for the determination of some of 
 the spin-flip
 amplitudes in the quasi-free charge-exchange scattering $pn\to np$
\cite{ellegaard,sams}. 

 The paper is organized in the following way. In section 2 we present the
general formalism for spin observables of a binary reaction in collinear
 kinematics.  
 In section 3 we present the full set of spin observables for the reaction 
 $\frac{1}{2}+\frac{1}{2}\to \frac{3}{2}+\frac{1}{2}$ and suggest a complete
 polarization experiment for this reaction.
 Section 4 deals with the reaction $\frac{1}{2}+1\to \frac{3}{2}+0$. 
 A complete polarization experiment for this reaction is described 
 in subsections 4.A and 4.B considering a transversally as well as 
longitudinally polarized beam. The relations between the amplitudes
 of the reactions $pd\to \Delta^0(pp)(^1S_0)$ and $ pn\to\Delta^0 p$ 
 are given in subsection 4.C in the impulse approximation.
 Within this approximation we find formulae for
 three independent linear combination of the amplitudes of
 the reaction  $ pn\to\Delta^0 p$ in terms of the invariant amplitudes
 of the reaction  $pd\to \Delta^0(pp)(^1S_0)$ and transition form factor
 \hbox{$d(^3S_1-^3D_1)\to^1S_0$}.
 In the Appendix formulae for 
 spin observables of the reaction $pd\to \Delta^0(pp)(^1S_0)$ are given.

\section{Formalism}
 The most general expression for the amplitude of the binary reaction
 $1+2 \to 3+4$ in collinear kinematics 
 can be written as \cite{GW}
\begin{eqnarray}
\label{tfi}
T_{\mu_1\,\mu_2}^{\mu_3\,\mu_4}= \langle \mu_1,\mu_2|F|\mu_3\mu_4 \rangle
= \sum_{S_i\,M_i \, S_f\,M_f\,L\,m}
(j_1\mu_1\,j_2 \mu_2|S_iM_i) \times \nonumber \\
\times (j_3\mu_3\, j_4\, \mu_4|S_f\,M_f)
 (S_i\,M_i\,L\,m|S_f\,M_f)
Y_{Lm}({\hat {\bf k}}) a_{S_f}^{LS_i} \ .
\end{eqnarray}
 Here $j_k$ and $\mu_k$ are the spin of the {\it k}-th particle and its  
 z-projection, $S_i(S_f)$ and $M_i(M_f)$ are the spin 
  and its z-projection for initial (final) particles. 
 The summation over the total angular momentum and orbital angular momenta
 in the initial and final states is included into the definition of the
 invariant spin amplitudes
 $a_{S_f}^{LS_i}$ (see Ref. \cite{uzikov3}). The orbital momentum $L$ in 
 Eq. (\ref{tfi}) is restricted by parity conservation via 
 $(-1)^L=\pi_1\pi_2\pi_3\pi_4$, where $\pi_i$ is the intrinsic
 parity of i-{\it th} particle.

 The triple-spin correlation coefficients for the case of two polarized 
 initial particles and one polarized final particle can be determined as 
\begin{equation}
\label{ttt}
K_{J_1\,M_1,J_2\, M_2}^{J_3\,M_3}=\frac{Tr\left \{
 T_{J_3M_3}(3)F\,
T_{J_1M_1}(1)\,T_{J_2M_2}(2)\,F^+\right\} }{TrFF^+},
\end{equation}
where $F$ is the transition operator determined by Eq. (\ref{tfi}),
and $ T_{J_iM_i}(i)$ denotes  
the tensor operator of rank $J_k$ and magnetic quantum number
$ M_k$  $(M=-J_k,-J_k+1, \cdots , J_k)$ for 
  the {\it i}-th particle. This operator is normalized as 
\begin{equation}
\label{normtt}
Tr T{JM}^+\,T_{J'M'}=\delta_{JJ'}\delta_{M'M}.
\end{equation}
Using Eqs.(\ref{tfi}), (\ref{ttt}) and properties of the operators $T_{JM}$ 
(see, for example, Ref. \cite{Varschalovich}),
one  can find the following formula \cite{uzikov3}
\begin{eqnarray}
\label{ktttgen}
K_{J_1\,M_1,J_2\, M_2}^{J_3\,M_3} \,Tr FF^+=
\frac{1}{4\pi}\,\sqrt{(2J_1+1)(2J_2+1)(2J_3+1)}\nonumber \\
\times \sum_{S\,S'\,J\,J'\,L\,L'\, J_0\,J_0'}
(2J+1)(2J'+1)\sqrt{(2L+1)\,(2L'+1)\,}\times\nonumber \\  
\times \sqrt{(2S+1)\,(2S'+1)\,(2J_0+1)}
(-1)^{j_3+j_4+J+L+S' - S}\times  \nonumber \\
\times (J_0\, -M_3 \, J_3\, M_3 |J_0'\, 0)
(L'\, 0 \, L\, 0\,|J_0' 0)(J_1M_1J_2M_2|J_0-M_3)
\times \ \ \  \nonumber \\
 \left \{\begin{array}{ccc}
j_3 & j_4 & J \\
J' & J_3 & j_3  
\end{array} \right \}
\left \{\begin{array}{ccc}
S & j_1 & j_2  \\
S' & j_1 & j_2  \\
J_0 & J_1 & J_2  
\end{array} \right \}
\left \{\begin{array}{ccc}
S & J & L \\
S' & J' & L' \\
J_0 & J_1 &  J_0' 
\end{array} \right \} a_J^{L\,S} (a_{J'}^{L'\,S'})^* .\ \ \ \ \ \  
\end{eqnarray}
 Due to the 
 presence of the Clebsch-Gordan coefficient $(L'0L0|J_0'0)$ in 
 Eq.~(\ref{ktttgen}) and parity conservation
 only even $J_0'$ contribute to the right side of
 Eq.~(\ref{ktttgen})
 \begin{eqnarray}
\label{evensum}
L+L'+ J_0' & {\text \,\,\, is\,\,\,  even,\,\,\,\,} &  
 J_0' {\text \,\,\ is \,\,\, even.}
\end{eqnarray}
 The coefficient $K_{J_1\,M_1,J_2\, M_2}^{J_3\,M_3}$ is nonzero only for
 $M_1+M_2+M_3=0$. From Eq.~(\ref{ktttgen}) one can find the following
property
\begin{equation}
\label{minusm}
K_{J_1\,-M_1, J_2\,-M_2}^{J_3\,-M_3}=(-1)^{J_1+J_2+J_3}
K_{J_1\,M_1, J_2\,M_2}^{J_3\,M_3}.
\end{equation}
 Eq. (\ref{minusm}) shows that, in the particular case of $M_1=M_2=M_3=0$
 only an even sum of $J_1+J_2+J_3$ 
 is allowed for nonzero triple correlations.
 
 Eq. (\ref{ktttgen}) is rather general and
 describes also double spin correlation if the rank $J_k$ for one (initial or
 final) particle is zero and it gives 
 tensor polarizations (or analyzing powers) for $J_k>1$ if the ranks of the two 
 other particles are zero.

 \section{The reaction $NN \to\Delta N$}

 For the reaction $NN \to\Delta N$ in collinear kinematics
 only four invariant amplitudes
 $a_{S_f}^{L\,S_i}$ are allowed by parity and angular momentum conservation:
 \begin{eqnarray}
\label{bbbb}
 B_1=a_2^{2\,0}, B_2= a_2^{2\,1}, B_3=a_1^{2\,1}, B_4=a_1^{0\,1},
\end{eqnarray}
 where $B_1$ results from the initial spin-singlet state and the other
 amplitudes 
 are from the spin-triplet state.

\subsection{Spin observables}

The unpolarized cross section can be written as
\begin{eqnarray}
\label{dsigm0}
d\sigma_0=\frac{\Phi}{(2j_1+1)(2j_2+1)}Tr{FF^+},
\end{eqnarray}
where
\begin{eqnarray}
\label{spur2}
Tr{FF^+}=\frac{1}{4\pi}\sum_{S\,S'\,L}(2S'+1)|a_{S'}^{LS}|^2
 \end{eqnarray}
and $\Phi$ is the phase-space factor.

 Using Eq.~(\ref{ktttgen}) one can find the full set of  spin observables
 of the reaction $NN \to\Delta N$  as 
\begin{eqnarray}
\label{k101000}
K_{10,10}^{00}\,\Sigma=
\frac{1}{4}(-5|B_1|^2+5|B_2|^2+|B_3+\sqrt{2}B_4|^2
 -|\sqrt{2}B_3-B_4|^2),\\
\label{k111100}
K_{11,1-1}^{00}\,\Sigma=\frac{1}{4}(5|B_1|^2- |\sqrt{2}B_3-B_4|^2),\\
\label{k100010}
K_{10,00}^{10}\,\Sigma=\frac{1}{8}\bigl [3\sqrt{5}|B_2|^2
+\sqrt{5}|B_3+\sqrt{2}B_4|^2 +2\sqrt{3}Re B_2^*(B_3+\sqrt{2}B_4)
-4Re B_1^*(\sqrt{2}B_3-B_4)\bigr ],\\
\label{k000020}
K_{00,00}^{20}\Sigma= \frac{1}{8}\Bigl [-10|B_1|^2-5|B_2|^2+2\sqrt{15}
Re B_2^*(B_3+\sqrt{2}B_4)
-2|\sqrt{2}B_3-B_4|^2+|B_3+\sqrt{2}B_4|^2\Bigr ],\\
\label{k101020}
K_{10,10}^{20}\Sigma= \frac{1}{8}\Bigl [10|B_1|^2-5|B_2|^2+2\sqrt{15}
Re B_2^*(B_3+\sqrt{2}B_4)
+2|\sqrt{2}B_3-B_4|^2+|B_3+\sqrt{2}B_4|^2\Bigr ].
\end{eqnarray}

\begin{eqnarray}
\label{k111120}
K_{11,1-1}^{20}=-K_{11,1-1}^{00},\\
\label{k111122}
K_{11,11}^{2-2}\Sigma= \frac{1}{8}\Bigl [-5\sqrt{6}|B_2|^2
-2\sqrt{10}Re B_2^*(B_3+\sqrt{2}B_4)
+\sqrt{6}|B_3+\sqrt{2}B_4|^2 \Bigr ],\\
\label{k100030}
K_{10,00}^{30}\Sigma= \frac{1}{4}\Bigl [-2\sqrt{5}|B_2|^2
 +6Re B_1^*(\sqrt{2}B_3-B_4)+2\sqrt{3}Re B_2^*(B_3+\sqrt{2}B_4)
 \Bigr ],\\
\label{k110031}
K_{11,00}^{3-1}\Sigma= \frac{1}{4}\Bigl [-2\sqrt{5}Re B_1\,B_2^*+
 4Re B_2^*(\sqrt{2}B_3-B_4)+2\sqrt{3}Re B_1^*(B_3+\sqrt{2}B_4)
 \Bigr ],\\
\label{sigma}
\Sigma=
5|B_1|^2+5|B_2|^2+|B_3+\sqrt{2}B_4|^2
 +|\sqrt{2}B_3-B_4|^2).
\end{eqnarray}

 For observables with the sum $J_1+J_2+J_3$ being odd (i.e. T-odd observables)
 we find the following formulae:
\begin{eqnarray}
\label{k11113-2}
K_{11,11}^{3-2}\,\Sigma=
\frac{i5\sqrt{2}}{2}Im B_2(B_3+\sqrt{2}B_4)^*,\\
\label{k111130}
K_{1-1,11}^{30}\,\Sigma=
\frac{i3}{2}Im B_1(\sqrt{2}B_3-B_4)^*,\\
\label{k111111}
K_{11,1-1}^{11}=K_{1-1,11}^{30}\,\\
\label{k111031}
K_{11,10}^{3-1}\,\Sigma=
\frac{i}{2}\Bigl [\sqrt{5}ImB_1B_2^*
-\sqrt{3}Im B_1(B_3+\sqrt{2}B_4)^*
+2Im B_2(\sqrt{2}B_3-B_4)^* \Bigl],\\
\label{k110021}
K_{11,00}^{2-1}\,\Sigma=
\frac{i}{16}\Bigl [-10\sqrt{2}ImB_1B_2^*
+2\sqrt{6}Im(B_3+\sqrt{2}B_4)(\sqrt{2}B_3-B_4)^*
-2\sqrt{30}Im B_1(B_3+\sqrt{2}B_4)^*+\nonumber \\ 
+2\sqrt{10}Im B_2(\sqrt{2}B_3-B_4)^* \Bigl].
\end{eqnarray}
One can see that the amplitudes $B_3$ and $B_4$ enter the
above formulae for observables only in the two combinations
$B_3+\sqrt{2}B_4$ and $\sqrt{2}B_3-B_4$.

\subsection{Complete polarization experiment}
 In order to determine  completely the matrix element of this reaction
 one has to find four moduli of the amplitudes and three relative phases
 (the overall phase is arbitrary). The choice of a minimal set of experiments
 is not unique and depends on the experimental conditions.  Here we describe 
 one possible minimal set.

 We use here and below the following relations:
\begin{eqnarray}
\label{modulphasa}
Re\, a_1\,a_2^*= |a_1||a_2|cos{(\phi_{a_1}-\phi_{a_2})},\,\,\,\,\,\,\,\,
Im\, a_1\,a_2^*= |a_1||a_2|sin{(\phi_{a_1}-\phi_{a_2})},
\end{eqnarray}
where $\phi_{a_i}$ is the phase of the amplitude ${a_i}$ ($i=1,2$).
The four moduli $|B_1|^2,\, |B_2|^2, \,|B_3+\sqrt{2}B_4|^2$ and
 $|\sqrt{2}B_3-B_4|^2)$ can be found from Eqs.
 (\ref{k111100}),
(\ref{k101020}), (\ref{k000020}) and (\ref{k111122}) in the following form 
 \begin{eqnarray}
\label{b1sqr2}
|B_1|^2=\Bigl \{ \frac{1}{5}\Bigl (K_{10,10}^{20}-K_{00,00}^{20}\Bigr )
+\frac{2}{5} K_{11,1-1}^{00}\Bigr\}\Sigma, \\ 
\label{b2sqr2}
|B_2|^2= \frac{1}{5}\Bigr \{1+K_{00,00}^{20}-\sqrt{6}
K_{11,11}^{2-2}-
3K_{10,10}^{20}\Bigr \}\Sigma, \\ 
\label{b34minus}
|\sqrt{2}B_3-B_4|^2=\Bigl \{ K_{10,10}^{20}-K_{00,00}^{20}-
2K_{11,1-1}^{00}\Bigr \}\Sigma, \\
\label{b34plus} 
|B_3+\sqrt{2}B_4|^2= \frac{1}{2}\left \{1+3K_{00,00}^{20}-
K_{10,10}^{20}+
\sqrt{6}K_{11,11}^{2-2}\right \}\Sigma,
\end{eqnarray}
where $\Sigma$ is given by Eq.(\ref{sigma}) and can be written as
 \begin{eqnarray}
\label{sigmadsg0}
\Sigma=4\pi Tr\,FF^+=\frac{16\pi d\sigma_0}{\Phi}.
\end{eqnarray}
 Thus, a measurement
 of the observables $d\sigma_0$, $K_{11,1-1}^{00}$,$K_{10,10}^{20}$,
 $K_{11,11}^{2-2}$ and $K_{00,00}^{20}$ is sufficient for 
 determining those moduli.
 Simultaneously, one can find
 from these observables the value of $Re B_2^*(B_3+\sqrt{2}B_4)$.
 In order to determine $Re B_1^*(\sqrt{2}B_3-B_4)$ one 
could measure $K_{10,00}^{30}$ in addition to the above observables.
 As a result, 
 for the two real parts we find the following
 \begin{eqnarray}
\label{realnnnd1}
Re B_2(B_3+\sqrt{2}B_4)^* =\frac{\Sigma}{2\sqrt{10}}
\Bigl \{ \sqrt{6}(K_{10,10}^{20}+K_{00,00}^{20})-2K_{11,11}^{2-2}
\Bigr \},\\
\label{realnnnd2}
Re B_1(\sqrt{2}B_3-B_4)^* ={\Sigma}\frac{1}{30}
\Bigl \{ \sqrt{5}+20K_{10,00}^{30}-2\sqrt{5}
K_{00,00}^{20}+3K_{10,10}^{20})
\Bigr \}.
\end{eqnarray}
 Furthermore, a measurement of the two T-odd observables
 $K_{11,11}^{3-2}$ and $K_{1-1,11}^{30}$ allows us to determine 
 $Im B_2(B_3+\sqrt{2}B_4)^*$ and $Im B_1(\sqrt{2}B_3-B_4)^*$, respectively,
 as it seen from Eqs. (\ref{k11113-2}) and (\ref{k111130}).
 Note that if the modulus and the real part of a complex number is known, 
then it would be sufficient to measure only the sign of the
 imaginary part in order to determine
 the phase of this number. Thus, measurement of
 the T-odd observables will
 not require a high accuracy.
 The knowledge of the real and imaginary parts of the products 
 $B_1(\sqrt{2}B_3-B_4)^*$ and $B_2(B_3+\sqrt{2}B_4)^*$
 completely determines the relative
 phases   $\phi_{B_1}-\phi_{\sqrt{2}B_3-B_4}$   and 
 $\phi_{B_2}-\phi_{B_3+\sqrt{2}B_4}$. 
The last relative phase, for example, 
$\kappa=\phi_{B_2}-\phi_{B_3+\sqrt{2}B_4}$,
can be found by a measurement of the observables $K_{11,10}^{3-1}$ and
$K_{11,00}^{3-1}$.
 Indeed, one can see from Eqs.~(\ref{k111031}) and (\ref{k110031}),
 that this measurement provides two linear equations for the two unknown
 variables $\sin{\kappa}$ and $\cos{\kappa}$. 

 Thus, a complete polarization experiment in this version requires to measure
 ten observables: $d\sigma_0$, $K_{11,1-1}^{00}$,$K_{10,10}^{20}$,
 $K_{11,11}^{2-2}$, $K_{00,00}^{20}$, $K_{10,00}^{30}$, $K_{11,00}^{31}$,
 $K_{11,11}^{3-2}$, $K_{1-1,11}^{30}$ and $ K_{11,10}^{3-1}$.

 \section{The reaction $pd \to \Delta^0 (pp)(^1S_0)$}

 For the reaction $pd \to\Delta^0 (pp)(^1S_0)$ in collinear kinematics
 the following three invariant amplitudes
 $a_{S_f}^{L\,S_i}$ are allowed by parity and angular momentum conservation:
 \begin{eqnarray}
\label{AAA}
 A_1=a_{\frac{3}{2}}^{2\,\frac{1}{2}},
 A_2=a_{\frac{3}{2}}^{0\,\frac{1}{2}}, 
 A_3=a_{\frac{3}{2}}^{2\,\frac{1}{2}},
\end{eqnarray}

 In order to determine a strategy for a complete polarization experiment we
 derive here all non-zero double and triple spin observables, which are given 
 in Appendix A. (Note that some observables are not presented but can be
 easily derived via Eq. (\ref{minusm}).) 
 In the subsequent discussion of a complete polarization experiment we 
 consider a particular case, namely those observables which
 require only transversally
 polarized particles in the initial state. The case with longitudinal
 polarizations is considered below separately. 

\subsection{Determination of the matrix element in measurements with
 transversally polarized beam and target}

 Using the following six observables, $d\sigma_0$, $K_{00,20}^{00}$,
$K_{1-1,11}^{00}$, $K_{00,00}^{20}$, $K_{00,21}^{2-1}$,
 and $K_{00,2-2}^{22}$,  
one can find moduli of the three amplitudes $|A_1|^2$, $|A_2|^2$, $|A_3|^2$
and  cosines of two phases:
\begin{eqnarray}
\label{fourpd}
|A_1|^2=I_0+\frac{2E+F}{3},\\ \nonumber
|A_2|^2=\frac{1}{2}C-\frac{1}{3}E-\frac{1}{6}F,\\ \nonumber
|A_3|^2=-\frac{1}{2}C-\frac{1}{3}E-\frac{1}{6}F,\\ \nonumber
Re A_1A_2^*= \frac{1}{2}D+\frac{E-F}{6},\\ \nonumber 
Re A_1A_3^*= \frac{1}{2}D-\frac{E-F}{6}, 
 \end{eqnarray}
where 
\begin{eqnarray}
\label{I0}
I_0=|A_1|^2+|A_2|^2+|A_3|^2=\pi TrFF^+=\frac{6\pi d\sigma_0}{\Phi},
\end{eqnarray}
\begin{eqnarray}
\label{fourpdconst}
E=\Bigl\{
4K_{1-1,11}^{00}-\frac{4}{\sqrt{3}}K_{00,20}^{00}-\frac{2}{3}\Bigr \}
{I_0},
\\ \nonumber
F=\Bigl\{
\frac{12}{\sqrt{3}}K_{00,20}^{00}-2{\sqrt{6}}K_{00,00}^{20}-1 \Bigr \}
{I_0},
\\ \nonumber
C=\Bigl\{
\frac{4}{\sqrt{3}}K_{00,2-2}^{22}-\frac{8}{\sqrt{3}}K_{00,21}^{2-1}
\Bigr \}
{I_0},
\\ \nonumber
D=-\frac{4}{\sqrt{3}}\Bigl\{
K_{00,2-2}^{22}+K_{00,21}^{2-1}\Bigr \}{I_0}.
 \end{eqnarray}
In order to find the sines of these phases one should measure the three
T-odd observables $K_{00,11}^{2-1}$, $K_{11,00}^{2-1}$, and  
$K_{11,2-1}^{00}$. With those observables one obtains
\begin{eqnarray}
\label{2todd}
iIm A_1A_2^*=-2\Bigl \{K_{11,2-1}^{00}-\frac{1}{\sqrt{3}}K_{00,11}^{2-1}+
\frac{1}{\sqrt{2}}K_{11,00}^{2-1}\Bigr \}{I_0},\\ \nonumber
iIm A_1A_3^*=2\Bigl \{K_{11,2-1}^{00}+\frac{1}{\sqrt{3}}K_{00,11}^{2-1}-
\frac{1}{\sqrt{2}}K_{11,00}^{2-1}\Bigr \}{I_0},\\ \nonumber
\end{eqnarray}
Note that for the sum $J_1+J_2+J_3$ being odd, the value of 
$K_{J_1M_1,J_2M_2}^{J_3M_2}$
 is purely imaginary.  The Cartesian components
 of these coefficients are purely real \cite{uzikov3}.
 Again, only the signs of the imaginary parts of $Im A_1A_2^*$ and
$Im A_1A_3^*$ are required.  
 Thus, one needs nine observables related to transversally polarized beam
 and/or target in order to completely determine the  three spin amplitudes 
 of the reaction $pd \to \Delta^0 (pp)(^1S_0)$.

\subsection{Determination of the matrix element by measurements with
 transversally and longitudinally polarized beam and target}

 Measurements with
 longitudinally polarized beam/target allows one
 to diminish the number of observables in a complete polarization experiment
as compared to measurements with transversal polarizations.
 For example, using $I_0$ and the longitudinal spin correlation parameter
 $K_{10,10}^{00}$ from Eq.~(\ref{s101000}) together
 with the transversal observables  $K_{1-1,11}^{00}$, 
$K_{00,2-1}^{21}$, $K_{00,2-2}^{22}$, given by Eqs. (\ref{I0}),
 (\ref{s111100}),
(\ref{s002121}) and (\ref{s002222}), respectively,
 one can find three moduli $|A_1|^2$, $|A_2|^2$, and
$|A_3|^2$. In addition, a measurement of the coefficient $K_{00,00}^{20}$
 given by Eq. (\ref{s000020}) allows to determine the cosines of three phases 
from $Re\, A_1A_2^*$, $Re\, A_1A_3^*$ and $Re\, A_2A_3^*$.
Finally, a knowledge of the two T-odd observables $K_{00,11}^{2-1}$ and
 $K_{11,00}^{2-1}$,
 presented by Eqs.~(\ref{s001121}) and (\ref{s110021}), 
gives $Im\, A_2A_3^*$ as
 \begin{eqnarray}
\label{1todd}
iIm A_2A_3^*=\Bigl \{
\frac{4}{\sqrt{3}}K_{00,11}^{2-1}+
\frac{2}{\sqrt{2}}K_{11,00}^{2-1}\Bigr \}{I_0}.\\ \nonumber
\end{eqnarray}
In view of the following relation between the relative phases,
 $\phi_{23}=\phi_{13}+\phi_{21}$,
where $\phi_{ij}=\phi_i-\phi_j$ and $\phi_i$ is  the  phase of the
amplitude $A_i\, (i=1,2,3)$, one can determine unambiguously the three 
complex numbers $A_1$, $A_2$ and $A_3$ from the above eight observables.  

\subsection{Impulse approximation for the transition amplitude}

\begin{figure}[hbt]
\includegraphics{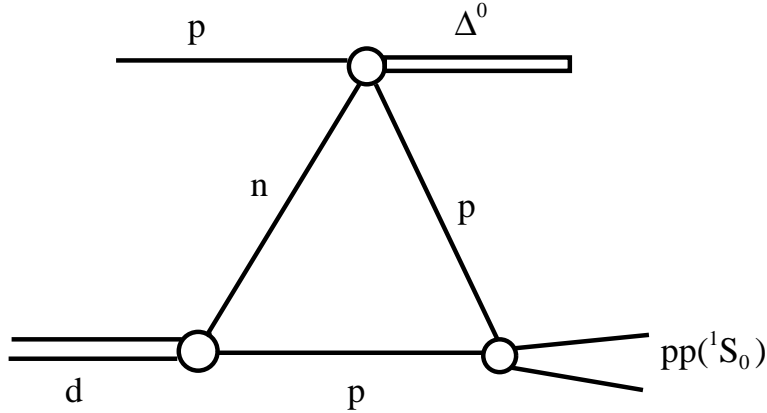}
\caption{\label{mechanism}The single scattering mechanisms of
 the $pd\to \Delta^0(pp)(^1S_0)n$ reaction.
}
\end{figure}

  Assuming that single $pN$ scattering (see Fig.~\ref{mechanism})
 dominates in the reaction 
 $pd \to \Delta^0 (pp)(^1S_0)$ one can express the transition amplitude
 of this reaction in terms of invariant amplitudes of the elementary
 subprocess $pn\to \Delta^0 p$ in the following way:
\begin{eqnarray}
\label{IAampl}
M_{\mu_0,\lambda}^{\mu_\Delta}(pd \to \Delta^0 (pp)_{^1S_0})=
-2\sqrt{m_N}\sum_{\mu_p\,\mu_n} (\frac{1}{2}\mu_p\frac{1}{2}-\mu_p|00)\times
\nonumber\\
\times \sum_{SLJ}(\frac{1}{2}\mu_0\frac{1}{2}\mu_n|SM_J)
(\frac{1}{2}-\mu_p\frac{3}{2}\mu_\Delta|JM_J)
(SM_JL0|JM_J)\times \nonumber \\
\times \sum_l(\frac{1}{2}\mu_p \frac{1}{2}\mu_n|1\mu_p+\mu_n )
(l01\mu_p+\mu_n|1\lambda)(-i)^l \frac{2l+1}{4\pi}
\sqrt{\frac{2L+1}{4\pi}}
S_l(Q/2) a_J^{LS} (pn\to \Delta^0 p) \ .
\end{eqnarray}
Here $\mu_0,\, \lambda$ and $\mu_\Delta$ are the $z$-projections  of the spins
 of the beam proton, deuteron and $\Delta$-isobar, respectively, and $m_N$ denotes
 the nucleon mass. $S_l(Q/2)$ ($l$=0 and 2) are the transition form
 factors $d(^3S_1-^3D_1)\to ^1S_0$ at the momentum Q transferred from the
 beam proton to the final $\Delta$ isobar,
\begin{eqnarray}
\label{FF}
S_l(Q/2)=\int_0^\infty dr r^2 j_l(Qr/2)u_l(r){\psi_{k}^{(-)}}^*(r),
\end{eqnarray}
where $u_0$ and $u_2$ are the S- and D-components of the deuteron wave function
 normalized by 
\begin{eqnarray}
\label{norma}
\int_0^\infty dr r^2 [u_0(r)^2+u_2(r)^2]=1,
\end{eqnarray}
The $NN$ scattering wave function in the $^1S_0$ state 
   is normalized at $r\to \infty $ as
 \begin{eqnarray}
\label{norma2}
\psi_{k}^{(-)}(r) \to \frac{sin(kr+\delta)}{kr}.
\end{eqnarray}
 where $k$ is the momentum of the nucleon
 in the cms of the $NN$ system
 and $\delta$ is the $^1S_0$ phase shift.

 Using Eq. (\ref{tfi}) one can find the invariant amplitudes  (\ref{AAA})
 from  the following relations:
 \begin{eqnarray}
\label{A3andM}
A_1=\frac{\sqrt{6\pi}}{3}\Bigl ( M_{+,0}^{+1/2}-\sqrt{2} M_{+,-}^{-1/2}\Bigr ),
\\ \nonumber
A_2=\frac{\sqrt{6\pi}}{3}\Bigl (\sqrt{\frac{3}{2}} M_{+,+}^{+3/2}+
 M_{+,0}^{+1/2} + \frac{1}{\sqrt{2}} M_{+,-}^{-1/2}\Bigr ),
\\ \nonumber
A_3=\frac{\sqrt{6\pi}}{3}\Bigl (\sqrt{\frac{3}{2}} M_{+,+}^{+3/2}- M_{+,0}^{+1}
-\frac{1}{\sqrt{2}} M_{+,-}^{-1/2}\Bigr ),
\end{eqnarray}
 where the lower indices in $M_{\mu_p,\lambda}^{\mu_\Delta}$
 correspond to the proton ($2\mu_p=\pm 1$) and  deuteron 
($\lambda=+1,0,-1$)
 spin projections on the $z$-axis.
 The above amplitudes are related to the unpolarized  c.m.s. cross section
 by 
 \begin{eqnarray}
\label{dscms}
\frac{d\sigma}{d\Omega_{c.m.}}=\frac{1}{64 \pi^2\, s}
\frac{p_f}{p_i}
\frac{1}{6}\sum_{\sigma_p,\lambda,\lambda_\Delta}
|M_{\sigma_p,\lambda}^{\lambda_\Delta}|^2=
\frac{1}{64 \pi^2\, s}\frac{p_f}{p_i}
\frac{1}{6\pi}(|A_1|^2+|A_2|^2+|A_3|^2),
 \end{eqnarray}
where $s$ denotes the invariant mass of the $pd$ system and $p_i$ ($p_f$)
is the initial (final) momentum in the c.m.s. of the binary reaction
$pd\to \Delta^0 (pp)(^1S_0)$.

 Using Eqs.~(\ref{tfi}), (\ref{IAampl}) and  (\ref{A3andM}) one can find
\begin{eqnarray}
\label{ASB}
A_1=\frac{\sqrt{6}}{192\pi}\Bigl \{
6\sqrt{5}S_0\,B_1 -(\sqrt{30}S_0 +10\sqrt{3}S_2)B_2-(3\sqrt{2}S_0
-6\sqrt{5}S_2)B_3 -6\sqrt{10}S_2B_4\Bigl \},  \nonumber \\
A_2=\frac{\sqrt{6}}{64\pi}\Bigl \{
5\sqrt{2}S_2\,B_1 -5\sqrt{3}S_2\,B_2+\sqrt{5}S_2B_3 -4S_0B_4\Bigl \},
  \nonumber \\
A_3=\frac{\sqrt{6}}{192\pi}\Bigl \{
15\sqrt{2}S_2\,B_1 -(2\sqrt{30}S_0 +5\sqrt{3}S_2)B_2+
(6\sqrt{2}S_0
-3\sqrt{5}S_2)B_3 -6\sqrt{10}S_2B_4\Bigl \}. 
\end{eqnarray}
 Solving the system of equations Eq.~(\ref{ASB}), one can find finally
 the following combinations of four amplitudes of the process
 $pn\to\Delta^0p$
\begin{eqnarray}
\label{3combin}
B_4=-\frac{16\pi}{\sqrt{15}}\Bigl \{(5S_2+\sqrt{10}S_0)A_2
+5S_2(A_3-\sqrt{3}A_1)\Bigl \}/Y,  \nonumber \\
\sqrt{5}B_2-\sqrt{3}B_3=
\frac{32\pi}{\sqrt{2}} \Bigl \{\sqrt{5}S_2(A_1+A_2)+
(\sqrt{2}S_0+\sqrt{5}S_2)A_3\Bigl \}/Y,  \nonumber \\
\sqrt{3}B_1-\sqrt{2}B_2=
\frac{16\pi}{\sqrt{5}} \Bigl \{ (2\sqrt{2}S_0+\sqrt{5}S_2)A_1
-3\sqrt{5}S_2A_2-(\sqrt{2}S_0-\sqrt{5}S_2)A_3\Bigr \}/Y,
\end{eqnarray}
where 
\begin{eqnarray}
\label{Y}
Y=2S_0^2+\sqrt{10}S_0S_2-10S_2^2.
\end{eqnarray}
 As one can see, the amplitudes $B_4$, $\sqrt{5}B_2-\sqrt{3}B_3$,
 and $\sqrt{3}B_1-\sqrt{2}B_2$ are determined by the three invariant amplitudes
 $A_1,\,A_2,\,A_3$, the form factor $S_0$ and the ratio $r=S_2/S_0$.
 The amplitudes $A_1,\,A_2,\,A_3$ can be measured by the complete polarization 
 experiment described by Eqs. (\ref{fourpd})-(\ref{2todd}). The transition
 form factor $S_0$ and the ratio $S_2/S_0$ are reasonably well constrained by  
existing $NN$ data at moderate transferred momenta $Q<300$ MeV/c
 (corresponding 
to the kinetic energy of the proton beam of $T_p> 800$ MeV) 

\section{ Concluding remarks}

  We have developed a general formalism for double and triple
 spin-correlations 
 of the reactions $NN\to \Delta N$ and $pd\to \Delta^0 (pp)(~^1S_0)$ in
 collinear kinematics in terms of, respectively,  four and three independent
 spin amplitudes, describing these reactions.
 A complete polarization experiment is suggested for
 the reaction $NN\to \Delta N$. 
 One possible set of observables is found in Eqs.~(\ref{b1sqr2}) -
 (\ref{realnnnd2}), (\ref{k110031}), (\ref{k11113-2}),
 (\ref{k111130}) and (\ref{k111031})
 which  includes seven T-even 
 observables and three T-odd ones. 
 For the reaction $pd\to \Delta^0 (pp)(~^1S_0)$ a complete 
 polarization experiment is described in Eqs. (\ref{fourpd}) - (\ref{2todd}) 
 in terms of nine observables related to 
 transversally polarized initial particles.
 We showed also that longitudinal observables used in combination with 
 transversal ones could reduce the total number of required measurements 
for a complete polarization experiment.
 On the basis of the impulse
 approximation for the reaction $pd\to \Delta^0 (pp)(~^1S_0)$,
 three combinations of the invariant amplitudes of the process
 $pn\to \Delta^0 p$ are expressed in Eqs.~(\ref{3combin}) and (\ref{Y})
 in terms of the invariant amplitudes
 of the reaction
 $pd\to \Delta^0 (pp)(~^1S_0)$.
 The formalism can be applied to the planned spin-physics at COSY
 \cite{spinCOSY}. 
In particular, the applicability of the impulse approximation
 for the reaction $pd\to \Delta^0 (pp)(~^1S_0)$ is expected to be valid
 at beam energies sufficiently high above the threshold, i.e. for $T>1$ GeV, 
 because then the transferred momentum is with $Q< 250$ MeV/c fairly small.
 Several spin observables which are necessary to perform the complete
 polarization experiments for the reactions $NN\to\Delta N$ and $pd\to \Delta^0
 (pp)({~^1S_0})$, require a measurement of the polarization of the final
 $\Delta$. Such a measurement can be performed
 by measuring the angular distribution of the final particles in the decay
 $\Delta\to \pi+N$,
 which are determined by the spin-density matrix of the $\Delta$ isobar
\cite{bermanjacob}. 
 Some of the spin-density matrix elements and single spin-correlations,  
 measured in the reaction
 ${\vec p}p\to \Delta^{++}n$ at beam momenta 3-11.8 GeV/c, were presented 
 in Ref. \cite{wicklund}.
 Besides of physics related to the $\Delta$ excitation, obviously this
 formalism can be applied to other baryon resonances with the spin-parity 
 $J^{\pi}=\frac{3}{2}^+$, in particular, to reactions involving the
 production of the $\Sigma^*$(1385) baryon.

\acknowledgments{We are thankful to C.~Wilkin who initiated this work. 
J.H. acknowledges instructive discussions with A. Nogga. 
Yu.U. thanks the Institut f\"ur Kernphysik at the Forschungzentrum J\"ulich 
for its hospitality during the period where part of this work was done. 
This work was supported in part by the BMBF grant WTZ Kaz-01/02 and the
Heisenberg-Landau program. 
}

\section*{Appendix A}
\setcounter{equation}{0}
\renewcommand{\theequation}{A.\arabic{equation}}
 Here we present the full set of nonzero spin observables for 
 the reaction $pd \to \Delta^0 (pp)(^1S_0)$
 in collinear kinematics. Using $I_0$ determined by
 Eq. (\ref{I0}), one can find from Eq. (\ref{ktttgen})
\begin{eqnarray}
\label{s111100}
K_{1-1,11}^{00}\, I_0=\frac{1}{24}\Bigl [ 4|A_1|^2-2|A_2|^2-2|A_3|^2
 + 2 Re(A_1A_2^*-A_1A_3^*+2A_2A_3^*\Bigr ],\\
\label{s002000}
K_{00,20}^{00}\, I_0=\frac{2\sqrt{3}}{12}\Bigl [
Re(A_1A_3^* -A_1A_2^*+A_2A_3^*)\Bigr ],\\
\label{s101000}
K_{10,10}^{00}\,I_0=-\frac{1}{12}\Bigl [ 2|A_1|^2-|A_2|^2-|A_3|^2
 - 2 Re(A_1A_2^*-A_1A_3^*+2A_2A_3^*)\Bigr ],\\
\label{s100010}
K_{10,00}^{10} \,I_0=-\frac{\sqrt{30}}{180}\Bigl
[|A_1|^2-5|A_2|^2-5|A_3|^2
-4Re(A_1A_2^*- A_1A_3^*+2A_2A_3^*)\Bigr ],\\
\label{s001010}
K_{00,10}^{10} \,I_0=\frac{\sqrt{5}}{60}\Bigl
[2|A_1|^2+5|A_2|^2+5|A_3|^2
+2Re(-A_1A_2^*+ A_1A_3^*+4A_2A_3^*)\Bigr ],\\
\label{s102010}
K_{10,20}^{10} \,I_0=-\frac{\sqrt{15}}{180}\Bigl
[4|A_1|^2-2|A_2|^2-2|A_3|^2
+2Re(A_1A_2^*- A_1A_3^*-7A_2A_3^*)\Bigr ],\\
\label{s1-1221-1}
K_{1-1,22}^{1-1}\, I_0=\frac{\sqrt{10}}{60}\Bigl
[4|A_1|^2+|A_2|^2+|A_3|^2
+2Re(-2A_1A_2^*+ 2A_1A_3^*-A_2A_3^*)\Bigr ],\\
\label{s112-110}
K_{11,2-1}^{10} \,I_0=-\frac{\sqrt{5}}{60}\Bigl
[-2|A_1|^2+|A_2|^2+|A_3|^2
-Re(A_1A_2^*- A_1A_3^*+2A_2A_3^*)\Bigr ],\\
\label{s112011}
K_{11,20}^{1-1} \,I_0=\frac{\sqrt{15}}{180}\Bigl
[4|A_1|^2+|A_2|^2+7|A_3|^2
+Re(14A_1A_2^* -2A_1A_3^*-8A_2A_3^*)\Bigr ],\\
\label{s110011}
K_{1-1,00}^{11} \,I_0=-\frac{\sqrt{30}}{180}\Bigl
[2|A_1|^2+5|A_2|^2-|A_3|^2
-2Re(A_1A_2^* +5A_1A_3^*+2A_2A_3^*)\Bigr ],\\
\label{s001111}
K_{00,1-1}^{11} \,I_0=\frac{\sqrt{5}}{120}\Bigl
[8|A_1|^2-10|A_2|^2+2|A_3|^2
-2Re(A_1A_2^* +5A_1A_3^* -4A_2A_3^*)\Bigr ], \\
\label{s102111}
K_{10,21}^{1-1} \,I_0=-\frac{\sqrt{5}}{60}\Bigl
[4|A_1|^2+|A_2|^2-5|A_3|^2
+Re(5A_1A_2^*+ A_1A_3^*+4A_2A_3^*)\Bigr ],
 \end{eqnarray}
\begin{eqnarray}
\label{s000020}
K_{00,00}^{20} \,I_0=\frac{\sqrt{6}}{12}\Bigl
[-|A_1|^2+2Re A_2A_3^*\Bigr ],\\
\label{s002020}
K_{00,20}^{20} \, I_0=\frac{\sqrt{3}}{12}\Bigl
[|A_2|^2+|A_3|^2+2Re(A_1A_2^*- A_1A_3^*)\Bigr ],\\
\label{s101020}
K_{10,10}^{20} \, I_0=\frac{1}{6}\Bigl
[|A_1|^2+|A_2|^2+|A_3|^2
+Re(-A_1A_2^* + A_1A_3^*+A_2A_3^*)\Bigr ],\\
\label{s111120}
K_{1-1,11}^{20} \, I_0=-\frac{1}{12}\Bigl
[2|A_1|^2-|A_2|^2-|A_3|^2
+Re(A_1A_2^*- A_1A_3^*+2A_2A_3^*)\Bigr ],\\
\label{s002121}
K_{00,2-1}^{21} \, I_0=\frac{\sqrt{3}}{12}\Bigl
[-|A_2|^2+|A_3|^2
-Re(A_1A_2^* +A_1A_3^*)\Bigr ],\\
\label{s111021}
K_{11,10}^{2-1} \, I_0 =\frac{\sqrt{3}}{12}\Bigl
[-|A_2|^2+|A_3|^2
+2Re(A_1A_2^*+ A_1A_3^*)\Bigr ],\\
\label{s101121}
K_{10,11}^{2-1}  \, I_0=-\frac{\sqrt{3}}{12}\Bigl
[|A_2|^2-|A_3|^2
+Re(A_1A_2^*+ A_1A_3^*)\Bigr ],\\
\label{s002222}
K_{00,2-2}^{22}  \, I_0=-\frac{\sqrt{3}}{12}\Bigl
[-|A_2|^2+|A_3|^2+2Re(A_1A_2^* +A_1A_3^*)\Bigr ],\\
\label{s111122}
K_{11,11}^{2-2} \, I_0 =\frac{\sqrt{6}}{12}\Bigl
[|A_2|^2-|A_3|^2
+Re(A_1A_2^*+ A_1A_3^*)\Bigr ],\\
\label{s112130}
K_{11,2-1}^{30}  \, I_0=\frac{\sqrt{5}}{20}\Bigl
[-2|A_1|^2+|A_2|^2+|A_3|^2
-Re(A_1A_2^*- A_1A_3^*+2A_2A_3^*)\Bigr ],\\
 \label{s102030}
K_{10,20}^{30}  \, I_0=\frac{\sqrt{15}}{60}\Bigl
[4|A_1|^2+3|A_2|^2+3|A_3|^2
+2Re(A_1A_2^*- A_1A_3^*-2A_2A_3^*)\Bigr ],\\
 \label{s001030}
K_{00,10}^{30} \, I_0 =\frac{\sqrt{5}}{10}\Bigl
[-|A_1|^2
+Re(A_1A_2^*- A_1A_3^*+A_2A_3^*)\Bigr ],\\
 \label{s100030}
K_{10,00}^{30}  \, I_0=\frac{\sqrt{30}}{60}\Bigl
[|A_1|^2
-2Re(2A_1A_2^*- 2A_1A_3^*-A_2A_3^*)\Bigr ],\\
 \label{s001131}
K_{00,1-1}^{31} \, I_0 =-\frac{\sqrt{30}}{30}\Bigl
[|A_1|^2-|A_3|^2
+Re(A_1A_2^*+A_2A_3^*)\Bigr ],\\
 \label{s110031}
K_{1-1,00}^{31} \, I_0 =\frac{\sqrt{5}}{30}\Bigl
[|A_1|^2+2|A_3|^2
+2Re(2A_1A_2^*-A_2A_3^*)\Bigr ],
\end{eqnarray}
\begin{eqnarray}
\label{s1-12031}
K_{1-1,20}^{31} \, I_0 =-\frac{\sqrt{10}}{60}\Bigl
[2|A_1|^2+ 3|A_2|^2+|A_3|^2+
Re(2A_1A_2^*-6 A_1A_3^* - 4A_2A_3^*)\Bigr ],\\ 
\label{s102-131}
K_{10,2-1}^{31} \, I_0 =\frac{\sqrt{30}}{30}\Bigl
[|A_1|^2-|A_2|^2-
Re(A_1A_3^* - A_2A_3^*)\Bigr ],\\
\label{s1-12-132}
K_{1-1,2-1}^{32}  \, I_0=\frac{\sqrt{6}}{12}\Bigl
[|A_2|^2-|A_3|^2+Re(A_1A_2^*+A_1A_3^*)\Bigr ],\\
\label{s102-232}
K_{10,2-2}^{32}  \, I_0=\frac{\sqrt{3}}{12}\Bigl
[|A_2|^2-|A_3|^2-Re(A_1A_2^*+A_1A_3^*)\Bigr ],\\ 
 \label{s1-12-233}
K_{1-1,2-2}^{33}  \, I_0=-\frac{1}{4}\Bigl
[|A_2|^2+|A_3|^2+ 2Re(A_2A_3^*)\Bigr ].
\end{eqnarray}
For the T-odd observabes we find
\begin{eqnarray}
\label{s112100}
K_{11,2-1}^{00}  \, I_0 =\frac{i}{4}Im(-A_1A_2^* +A_1A_3^*),\\
\label{s111110}
K_{11,1-1}^{10}  \, I_0 =\frac{i\sqrt{5}}{20}Im(A_1A_3^* -A_1A_2^*), \\
\label{s111011}
K_{11,10}^{1-1}  \, I_0 =i\frac{\sqrt{5}}{10}
Im(A_2A_1^*+ A_3A_1^*+A_2A_3^*),\\
\label{s101111}
K_{10,1-1}^{11}  \, I_0 =-i\frac{\sqrt{5}}{20}
Im(A_2A_1^*+ 3A_1A_3^*+2A_2A_3^*),\\
\label{s002111}
K_{00,21}^{1-1}  \, I_0 =i\frac{\sqrt{5}}{20}
Im(3A_1A_2^*+ A_3A_1^*+2A_2A_3^*),\\
\label{s001121}
K_{00,11}^{2-1}  \, I_0 =\frac{i\sqrt{3}}{12}Im(2A_1A_2^* +A_1A_3^*
+2A_2A_3^*),\\ 
\label{s110021}
K_{11,00}^{2-1}  \, I_0 =-\frac{i\sqrt{2}}{6}Im(A_1A_2^* +A_1A_3^*
+A_3A_2^*),\\ 
\label{sk10212-1}
K_{10,21}^{2-1}  \, I_0 =i\frac{\sqrt{3}}{12}
Im(A_1A_2^*+ A_1A_3^*+2A_2A_3^*),\\
\label{sk11202-1}
K_{11,20}^{2-1} \, I_0  =i\frac{1}{6}
Im(A_1A_2^*+ A_1A_3^*-A_2A_3^*),\\
\label{sk112-120}
K_{11,2-1}^{20}  \, I_0 =i\frac{1}{4}
Im(A_1A_2^*-A_1A_3^*),\\
\label{sk112122}
K_{11,21}^{2-2} \, I_0  =i\frac{\sqrt{6}}{12}
Im(A_2A_1^*+ A_3A_1^*+2A_3A_2^*),\\
\label{s10222-2}
K_{10,22}^{2-2}  \, I_0 =i\frac{\sqrt{3}}{6}
Im(A_2A_1^*+ A_3A_1^*+A_2A_3^*),\\
\label{s101-131}
K_{10,1-1}^{31}  \, I_0 =-i\frac{\sqrt{30}}{30}
Im(2A_1A_2^*+ A_3A_1^*+A_2A_3^*),\\
\label{s002-232}
K_{00,2-2}^{32} \, I_0  =i\frac{\sqrt{3}}{6}
Im(A_2A_1^*+ A_3A_1^*+A_2A_3^*),\\
\label{s1-11031}
K_{1-1,10}^{31}=
K_{00,2-2}^{32}.
\end{eqnarray}
In addition, we have 
\begin{equation}
\label{zeros}
K_{1-1,22}^{2-1}=0.
\end{equation}



\end{document}